\documentclass{pasj01}
\draft

\usepackage{times}
\usepackage{lscape}
\usepackage{url}
\usepackage{color}
\usepackage{ulem}

\newcommand{\kt}{\ensuremath{k_{\rm{B}}T}}
\newcommand{\lx}{\ensuremath{L_{\rm{X}}}}

\newcommand{\fx}{\ensuremath{F_{\rm{X}}}}

\newcommand{\fks}{\ensuremath{F_{\rm{Ks}}}}

\newcommand{\nh}{\ensuremath{N_{\rm H}}}
\newcommand{\ksi}{\textit{K\ensuremath{_{\rm{s}}}}}

\newcommand{\ergcms}{{\rm ergs}\ {\rm cm}^{-2}\ {\rm s}^{-1}}

\newcommand{\hi}{\textit{H}}

\def\maru#1{\textcircled{\scriptsize#1}}

\begin{document}
\SetRunningHead{Morihana et al.}{NIR Spectroscopy of Faint X-ray Sources Constituting GRXE}
\Received{}
\Accepted{}
\Published{}

\title{Near-Infrared Spectroscopy of Faint Discrete X-ray Point Sources Constituting the Galactic Ridge X-ray Emission}

\author{Kumiko \textsc{Morihana}\altaffilmark{1}, Masahiro \textsc{Tsujimoto}\altaffilmark{2}, Pierre \textsc{Dubath}\altaffilmark{3}, Tessei \textsc{Yoshida}\altaffilmark{4}, 
Kensuke \textsc{Suzuki}\altaffilmark{5}, and Ken \textsc{Ebisawa}\altaffilmark{2}}

\altaffiltext{1}{Nishi-Harima Astronomical Observatory, Center for Astronomy, University of Hyogo, 407-2 Nichigaichi, Sayo-cho, Sayo, Hyogo, 670-5313}
\altaffiltext{2}{Japan Aerospace Exploration Agency, Institute of Space and Astronautical Science, 3-1-1 Yoshino-dai, Chuo-ku, Sagamihara, Kanagawa 252-5210}
\altaffiltext{3}{Department of Astronomy, University of Geneva, 16 Ch. d'Ecogia, 1290, Versoix, Switzerland}
\altaffiltext{4}{National Astronomical Observatory of Japan, 2-21-1 Osawa, Mitaka, Tokyo, 181-8588}
\altaffiltext{5}{TIS, 8-17-1 Nishi-shinjuku, Shinjuku-ku, Tokyo, 160-0023}

\email{morihana@nhao.jp}


\maketitle

\begin{abstract}
The Galactic Ridge X-ray Emission (GRXE) is apparently extended X-ray emission along the
Galactic Plane. The X-ray spectrum is characterized by hard continuum with a strong Fe K
emission feature in the 6--7 keV band. A substantial fraction ($\sim$80\%) of the GRXE
in the Fe band was resolved into point sources by deep Chandra imaging observations,
thus GRXE is mostly composed of dim Galactic X-ray point sources at least in this energy
band. To investigate the populations of these dim X-ray point sources, we carried out
near-infrared (NIR) follow-up spectroscopic observations in two deep Chandra fields
located in the Galactic plane at ($l$,$b$)$=$(0.1\arcdeg, --1.4\arcdeg) and
(28.5\arcdeg, 0.0\arcdeg) using NTT/SofI and Subaru/MOIRCS. We obtained well-exposed NIR
spectra from 65 objects and found that there are three main classes of Galactic
sources based on the X-ray color and NIR spectral features: those having (A) hard X-ray
spectra and NIR emission features such as H\emissiontype{I} (Br$\gamma$),
He\emissiontype{I}, and He\emissiontype{II} (2 objects), (B) soft X-ray spectra and NIR
absorption features such as H\emissiontype{I}, Na\emissiontype{I},
Ca\emissiontype{I}, and CO (46 objects), and (C) hard X-ray spectra and NIR
absorption features such as H\emissiontype{I}, Na\emissiontype{I}, Ca\emissiontype{I}
and CO (17 objects). From these features, we argue that class A sources are cataclysmic
variables (CVs), and class B sources are late-type stars with enhanced coronal activity,
which is in agreement with current knowledge. Class C sources  possibly belong to a new group
of objects, which has been poorly studied so far.
We argue that the candidate sources for class C are the binary systems hosting
white dwarfs and late-type companions with very low accretion rates.
It is likely that this newly recognized class of the sources contribute to a non-negligible fraction of the
 GRXE, especially in the Fe K band.
\end{abstract}

%
\section{Introduction}\label{s1}
Since the dawn of the X-ray astronomy, apparently diffuse emission of low surface
brightness has been known to exist along the Galactic Plane (GP; $|l|<$45$^{\circ}$,
$|b|<$1$^{\circ}$)), which is referred to as the Galactic Ridge X-ray Emission (GRXE;
e.g., \cite{worrall82, warwick85, koyama86}). The X-ray spectrum is characterized by hard
continuum with strong Fe K emission lines in the hard (2--8~keV) X-ray band
(e.g., \cite{koyama86}).
The origin of the GRXE has been a mystery for a long time. A long-standing debate was
whether it is a truly diffuse plasma \citep{ebisawa01,ebisawa05} or the sum of dim
unresolved X-ray point sources \citep{revnivtsev06}. \citet{revnivtsev09}
claimed that most of the GRXE was resolved into dim X-ray point sources down to
$\sim$10$^{-16}$$\ergcms$ in the hard band using the deepest observation ($\sim$900 ks) 
made with the Chandra X-ray Observatory \citep{weisskopf02} at a slightly off-plane region of
($l$,$b$)$=$(0.1\arcdeg, --1.4\arcdeg) in the Galactic bulge (Chandra bulge field; hereafter,
CBF), which also known as the Limiting Window in the studies of \citet{Berg09} and \citet{hong12}.

If the GRXE is composed of numerous dim X-ray point sources in the hard band, new
questions arise : \textit{What are the populations of these X-ray point sources? Which
class of sources contribute to the Fe K emission lines?} There are several candidates
for such sources. \citet{Revnivtsev11} claimed that the majority of faint sources
are X-ray active stars based on the luminosity function, while \citet{hong12} argued
that hard X-ray component of the GRXE above 3 keV is dominated by sources
such as magnetic cataclysmic variables (mCVs). However, it is difficult to constrain the
nature of these point sources from X-ray data alone, because most of these sources are
detected only with a limited number of X-ray photons even with the deepest
observations. Follow-up spectroscopic studies in longer wavelengths are keys to
elucidate the nature of individual X-ray sources.

In the optical band, several spectroscopic observations were carried out \citep{motch10,
van12, servillat12, Nebot13, Berg06}. They revealed that some X-ray sources are mCVs and
late-type active stars, which was indeed speculated from optical and near-infrared (NIR)
imaging studies of their work and other preceding work \citep{Muno04, Muno09, Berg09,
ebisawa05}. They also pointed out a small contribution from other classes of source,
such as young stellar objects, Wolf-Rayet stars, $\gamma$Cas analogues, and symbiotic
binaries.

Most of the X-ray sources studied in these optical studies are biased toward bright, or
close, X-ray sources. This is because (i) the two XMM-Newton studies \citep{motch10,
Nebot13} are based on wide and shallow surveys, which is $\sim$1 order magnitude
shallower than the Chandra deep Galactic plane survey such as \citet{ebisawa05} and
\citet{Grindley05}, and (ii) the majority of the discrete X-ray sources comprising the
GRXE are too attenuated by interstellar extinction to allow optical identification.
\citet{motch10} estimated that only CVs within a 2~kpc distance are accessible within
their optical survey limit.

To proceed further, we embarked on NIR spectroscopic follow-up studies based on the
deepest Chandra data. \ksi-band extinction is about 10 times lower than \textit{V}-band
extinction, allowing access to the sources behind a large column of extinction. Here, we
present the results of NIR follow-up study of dim X-ray sources constituting the
GRXE. The outline of this paper is as follows. In $\S$\ref{s2}, we present the data set
and reduction. In $\S$\ref{s3}, we show NIR spectra and classify sources based on the
NIR spectra and X-ray colors. In $\S$\ref{s4}, we discuss the nature of these classes and
their contributions to the GRXE. Finally, we summarize our results in $\S$\ref{s5}.

\section{Observations}\label{s2}
\subsection{Data Sets}\label{s2-1}
We first summarize the data sets in table~\ref{t01}. We use two GP fields that were
studied with deep Chandra exposures: the CBF and the Ebisawa field at ($l$,
$b$)$=$(28.5\arcdeg, 0.0\arcdeg) by \citet{ebisawa01}. These two fields were studied in
the NIR imaging surveys using the Simultaneous Infrared Imager for Unbiased Survey
(SIRIUS; \cite{nagashima99},~\cite{nagayama03}) on the Infrared Survey Facility (IRSF)
in the South African Astronomical Observatory and the Son of ISAAC (SofI:
\cite{moorwood98}) on the New Technology Telescope (NTT; \cite{tarenghi89}) in the
European Southern Observatory by \citet{morihana12} and \citet{ebisawa05},
respectively. In total, 222 out of 2002 (CBF) and 142 out of 274 (Ebisawa field) X-ray
sources were identified with NIR sources in these studies. 
\begin{table*}[htbp]
  \caption{Data sets}\label{t01}
  \begin{center}
  \begin{tabular}{llcc}
   \hline
   \hline
   Observations  & & CBF & Ebisawa field \\
   \hline
                    & Telescope/Instrument & Chandra/ACIS-I            & Chandra/ACIS-I \\
	            & Reference            & \citet{morihana13} & \citet{ebisawa05} \\
   \hline
   NIR imaging      & Number of sources\footnotemark[$\dagger$] & 222 & 142 \\
                    & Telescope/Instrument & IRSF/SIRIUS               & NTT/SofI \\
	            & Reference            & \citet{morihana12} & \citet{ebisawa05} \\
   \hline
   NIR spectroscopy & Number of sources\footnotemark[$\ddagger$]&  23 & 42\\
                    & Telescope/Instrument & Subaru/MOIRCS             & NTT/SofI $+$ Subaru/MOIRCS\\
	            & Reference            & (this study)       & (this study) \\
   \hline
   \multicolumn{4}{@{}l@{}}{\hbox to 0pt{\parbox{135mm}{
   \footnotesize
   \par \noindent
   \footnotemark[$*$]The number of detected X-ray point sources.\\
   \footnotemark[$\dagger$]The number of X-ray point sources identified in the NIR imaging studies.\\
   \footnotemark[$\ddagger$]The number of NIR-identified X-ray point sources, for which well-exposed NIR spectra were obtained in this study. \\
   }\hss}}
  \end{tabular}
 \end{center}
\end{table*}

We selected our NIR spectroscopy targets among these NIR-identified sources based on
their X-ray spectral hardness, variability, and brightness. 
We used the NTT (\S~\ref{s2-2-1}) and the Subaru Telescope (\S~\ref{s2-2-2})
for observations.
As our interest is to study
origin of the sources contributing to the hard-band emission of GRXE, the selected
targets are preferentially hard X-ray sources. Also, we selected sources with
$K_{\mathrm{s}}$-band magnitude brighter than 14~mag for a reasonable telescope
time. Among such sources, the false positive rate (a wrong counterpart pairs by chance)
is estimated to be $\sim$3\% (CBF) and $\sim$1\% (Ebisawa field) by the product of the
surface number density of NIR sources with $m_{K_{\mathrm{s}}}<14$~mag and the maximum
search circle. This is reasonably low to discuss statistical properties of subgroups,
but the identification of individual sources is subject to contamination by false
positives. 

\subsection{Observations \& Data Reduction}\label{s2-2}
\subsubsection{NTT/SofI}\label{s2-2-1}
We conducted a pilot observation of X-ray sources in the Ebisawa field using the SofI
mounted at the Nasmyth focus of the NTT, which is
located at the La Silla Observatory, Chile. The NTT is a 3.6~m Ritchey-Chr\'{e}tien
telescope on an altazimuth mount. The SofI is equipped with a $1024 \times 1024$ pixel
HgCdTe infrared array, and we used its long-slit spectroscopy mode with a
medium-resolution grism ($R\sim$ 2200) in the \ksi~band (2.0--2.3~$\mu$m) at a
dispersion scale of $\sim$4.62~\AA~pixel$^{-1}$.

\begin{table*}[htbp]
 \caption{NTT observation log}\label{t02}
\begin{center}
  \begin{tabular}{lcccccc}
   \hline
   \hline
   Ref.\footnotemark[]{$*$} & \multicolumn{2}{c}{Coordinate~(J2000.0)} &
SW\footnotemark[]{$\dagger$} &   $t_{\rm{exp}}$\footnotemark[]{$\S$} & Airmass 
& Seeing\footnotemark[]{$\ddagger$} \\
  \cline{2-3}
   ID   & R.\,A. & Decl.  & (arcsec) & (min) &  & (\arcsec)\\
   \hline
   SoE3    & 18:42:58.3 & $-$03:53:27 & 1.0 & 20 & 1.14--1.19 & 1.1 \\
   SoE4    & 18:43:00.4 & $-$03:53:49 & 1.0 & 20 & 1.11--1.11 & 1.3\\
   SoE29   & 18:43:17.5 & $-$03:56:00 & 1.0 & \phantom{0}4 & 1.11--1.12 & 1.0\\
   SoE70   & 18:43:28.4 & $-$04:07:33 & 1.0 & \phantom{0}2 & 1.68--1.76 & 1.2\\
   SoE79   & 18:43:29.7 & $-$03:50:15 & 1.0 & 20 & 1.12--1.16 & 1.8\\
   SoE86   & 18:43:30.8 & $-$04:01:03 & 1.0 & \phantom{0}2 & 1.91--2.01 & 1.1\\
   SoE100  & 18:43:32.6 & $-$04:04:19 & 1.0 & \phantom{0}4 & 1.38--1.45 & 0.9\\
   SoE104  & 18:43:33.5 & $-$04:03:54 & 2.0 & \phantom{0}2 & 1.54--1.60 & 1.2\\
   SoE105  & 18:43:33.9 & $-$03:52:53 & 1.0 & 36 & 1.16--1.28 & 1.9\\
   SoE135  & 18:43:39.2 & $-$03:52:53 & 1.0 & \phantom{0}2 & 2.06--2.17 & 1.2\\
   SoE221  & 18:43:59.7 & $-$03:55:18 & 1.0 & \phantom{0}2 & 1.78--1.87 & 1.1\\
   SoE233  & 18:44:03.9 & $-$04:02:58 & 1.0 & 36 & 1.16--1.29 & 1.1\\
   SoE244  & 18:44:11.0 & $-$04:05:17 & 1.0 & 24 & 1.10--1.13 & 1.4\\
   SoE255  & 18:44:18.6 & $-$04:06:03 & 1.0 & 20 & 1.11--1.13 & 1.8\\
   SoE272  & 18:44:28.8 & $-$04:01:03 & 2.0 & 16 & 1.62--1.92 & 1.8\\
   \hline
\multicolumn{4}{@{}l@{}}{\hbox to 0pt{\parbox{135mm}{
   \footnotesize
\par \noindent
\footnotemark[$*$]The reference  source numbers follow \citet{ebisawa05}.\\
\footnotemark[$\dagger$] Slit width.\\
\footnotemark[$\S$]  Total exposure time.\\
\footnotemark[$\ddagger$] Average seeing of all images in each slit.\\
 }\hss}}
\end{tabular}
\end{center}
\end{table*}

\begin{figure}
 \begin{center}
  \FigureFile(85mm,85mm){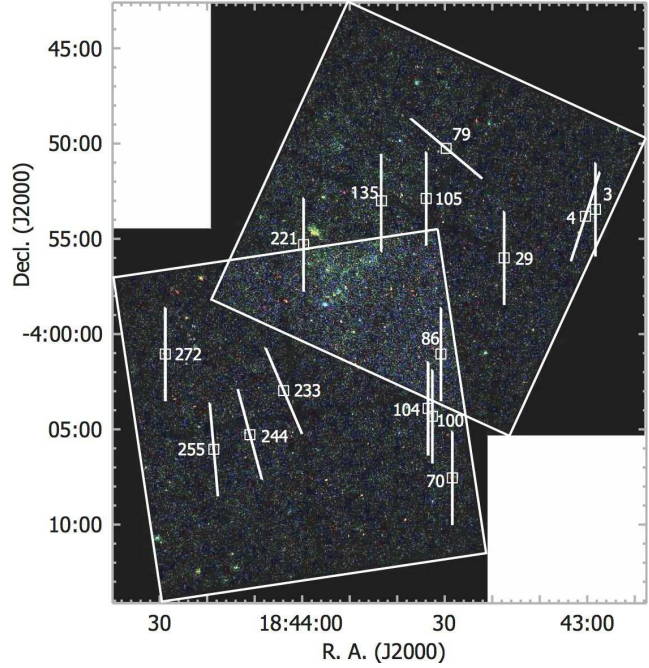}
 \end{center}
 \caption{Positions of the long slits for the SofI spectroscopy in the Ebisawa
 field. The background is the Chandra pseudo-color X-ray image and the boxes show the
 source location (see Table~\ref{t02}) used for the NIR spectroscopy with the sources
 numbers given in \citet{ebisawa05}.}\label{f01}
\end{figure}

We carried out the observation on 2005 July 18 and 19 for 16 selected X-ray sources, and
well-exposed spectra were obtained from 15 of them (figure~\ref{f01},
table~\ref{t02}). Standard stars were also observed for the telluric correction, and the
Xe-Ne lamp spectra were obtained for the wavelength calibration. Each spectrum was
integrated for 30 or 60~s using a 1\farcs0 or 2\farcs0 slit width depending on their
brightness. A set of spectra was composed of four spectra at two dithered positions (A
and B) that are 60\arcsec\ apart along the slit. From one to nine sets were obtained,
resulting in a total net exposure time of 2 to 36 minutes. The seeing was 0\farcs9 to
1\farcs9 on the first night and 1\farcs0 to 1\farcs2 on the second night.

The SofI data were reduced using the standard procedures with the IRAF software
package\footnote{http://iraf.noao.edu/}. The frames were flat-fielded using dome flats
and cleaned removing bad pixels and cosmic-ray events. Frames taken at the A and B
positions were subtracted from each other to remove the dark current and sky
emission. The frames were then combined, and the source spectra were extracted,
registered for the wavelengths using the comparison lamp spectra fitted by a third-order
polynomial. The calibration uncertainty is $\sim$0.32~\AA.  We then corrected for
telluric features by dividing the A0\,V standard star spectra, in which the Br$\gamma$
feature is removed by a local Voigt profile fitting and the continuum is flattened by a
9790~K blackbody emission.

\subsubsection{Subaru/MOIRCS}\label{s2-2-2}

\begin{figure}[htbp]
 \begin{center}
  \FigureFile(85mm,85mm){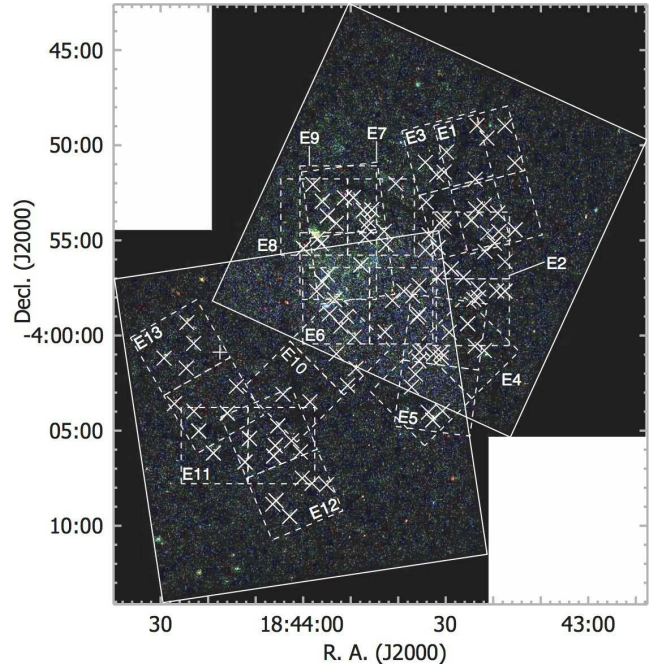}
 \end{center}
 \caption{Positions of the masks for the MOIRCS spectroscopy in the Ebisawa field. White
 dashed boxes show masks correspond to the Mask ID in table~\ref{t03}. White crosses
 indicate the source positions for which NIR spectra were obtained.}\label{f02}
\end{figure}

\begin{figure}[htbp]
 \begin{center}
  \FigureFile(85mm,85mm){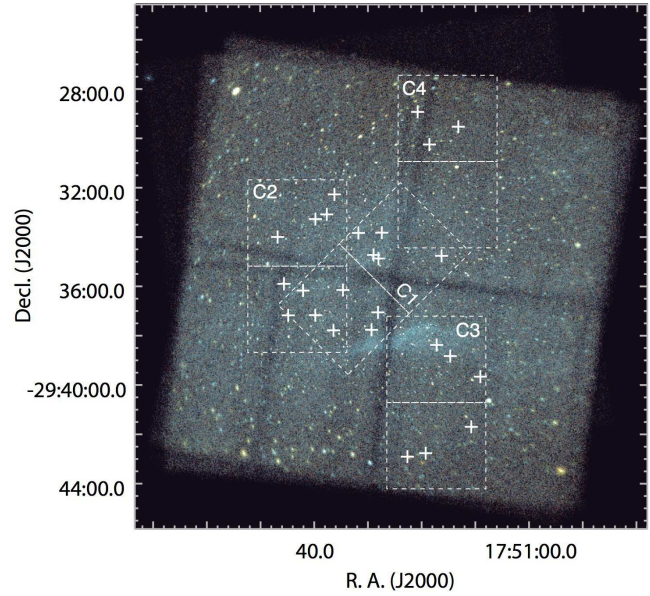}
 \end{center}
 \caption{Positions of the masks for the MOIRCS spectroscopy in the CBF. The symbols are
 the same as in figure~\ref{f02}.}
 \label{f03}
\end{figure}

\begin{figure*}[htbp]
 \begin{center}
  \FigureFile(150mm,60mm){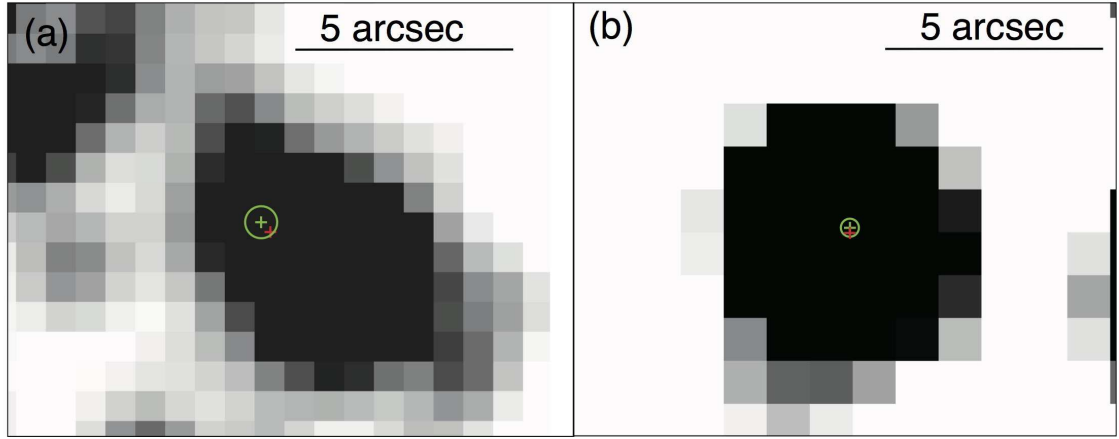}
 \end{center}
 \caption{Close-up view of the $K_{\mathrm{s}}$-band image around two class A sources
 (a) source number 79 and (b) 100. The X-ray source position and its uncertainty are
 shown with a green cross and circle, while the position of its NIR counterpart is shown
 with a red cross.}\label{f04}
\end{figure*}

\begin{table*}[htbp]
 \caption{Subaru observation log}\label{t03}
\begin{center}
 \begin{tabular}{lcccccc}
  \hline
   \hline
  Mask ID\footnotemark[]{$*$}  & \multicolumn{2}{c}{Coordinate~(J2000.0)} &
   $t_{\mathrm{exp}}$\footnotemark[]{$\dagger$}   &   Num. of slits & Airmass 
& Seeing\footnotemark[]{$\ddagger$} \\
  \cline{2-3}
                  & R.\,A. & Decl.     & (min) &  &  & (\arcsec) \\
   \hline
   C1    & 17:51:29 & $-$29:35:43 & 180 & 15 & 1.55--1.97 & 0.8\\
   C2    & 17:51:43 & $-$29:34:47 & 165 & 15 & 1.63--2.02 & 1.3\\
   C3    & 17:51:17 & $-$29:40:60 & 140 & 13 & 1.55--1.83 & 0.7\\
   C4    & 17:51:16 & $-$29:31:02 & 160 & 10 & 1.53--1.85 & 0.9\\
   \hline
   E1    & 18:42:52 & $-$03:53:29 &  \phantom{0}12 &\phantom{0}7  & 1.00--1.62 & 0.3\\
   E2    & 18:42:53 & $-$03:52:01 &  \phantom{0}24 &\phantom{0}7  & 1.09--1.29 & 0.3\\
   E3    & 18:43:24 & $-$04:02:01 &  \phantom{0}32 &14 & 1.47--1.73 & 0.5\\
   E4    & 18:43:50 & $-$03:53:40 &  \phantom{0}24 &14 & 1.23--1.88 & 0.5\\
   E5    & 18:43:30 & $-$04:01:26 &  \phantom{0}18 & \phantom{0}9 & 1.09--1.32 & 0.3\\
   E6    & 18:43:45 & $-$03:58:20 &  \phantom{0}24 &11 & 1.22--1.52 & 0.7\\
   E7    & 18:43:50 & $-$03:53:39 &  \phantom{0}30 &12 & 1.09--1.14 & 0.3\\
   E8    & 18:43:40 & $-$04:01:48 &  \phantom{0}45 & \phantom{0}7 & 1.21--1.60 & 0.3\\
   E9    & 18:43:25 & $-$03:56:55 &  \phantom{0}40 &11 & 1.09--1.12 & 0.4\\
   E10   & 18:43:57 & $-$04:01:37 &  \phantom{0}40 & \phantom{0}6 & 1.18--1.49 & 0.4\\
   E11   & 18:43:21 & $-$03:51:41 &  \phantom{0}45 & \phantom{0}9 & 1.21--1.68 & 0.4\\
   E12   & 18:43:51 & $-$03:54:20 &  \phantom{0}45 & \phantom{0}9 & 1.09--1.12 & 0.5\\
   E13   & 18:44:05 & $-$04:06:26 &  \phantom{0}40 &11 & 1.20--1.59 & 0.9\\
   \hline
\multicolumn{4}{@{}l@{}}{\hbox to 0pt{\parbox{135mm}{
   \footnotesize
\par \noindent
\footnotemark[$*$] ``C'' shows the CBF and ``E'' shows the Ebisawa field.\\
\footnotemark[$\dagger$]
Exposure time.\\
\footnotemark[$\ddagger$]
 Average seeing of all images in each mask.\\
   }\hss}}
 \end{tabular}
\end{center}
\end{table*}

In order to increase the number of samples and to extend the samples to the fainter end,
we further conducted multi-object NIR spectroscopy of X-ray sources both in the Ebisawa
field and the CBF. We used the Multi-Object Infra-Red Camera and Spectrograph (MOIRCS;
\cite{ichikawa06}) at the Cassegrain focus of the 8.2~m Subaru Telescope in the Mauna
Kea Observatory. MOIRCS is equipped with two adjacent HAWAII-2 arrays of
2048$\times$2048 pixels. The instrument is capable of obtaining multiple spectra
simultaneously using a mask with multiple slits designed specifically for each field. We
used the $R\sim$1300 grism in the \ksi-band (2.0--2.3~$\mu$m) with a dispersion scale
of $\sim$3.88~\AA~pixel$^{-1}$.

The observations were performed on 2007 June 07--10, 2008 June 27--28 (the Ebisawa
field; figure~\ref{f02}) and on 2011 June 23--24 (the CBF; figure \ref{f03}). We
designed masks so that we can cover as many sources as possible. Each mask has 6--15
slits (table~\ref{t03}), and 2--3 masks were used each night. The A--B--B--A dithering
pattern was repeated several times for each mask. We used the OH emission lines in the
sky spectra for the wavelength calibration in both fields. This resulted in saving
some telescope time, but a degraded calibration accuracy in the longer wavelengths in
the \ksi-band where the OH emission features are sparse. In each night, we took
standard stars several times using one of the slits of a mask, so that one of them has a
similar airmass with the target objects. 

The MOIRCS data were also reduced using the standard procedure in the same manner as the
SofI data. In addition, we applied the distortion correction according to the recipe
provided by the instrument team\footnote{See
http://www.naoj.org/staff/ichi/MCSRED/mcsred.html for detail.}.  We generated both A--B
and A+B frames respectively for the source and sky spectra; the latter were used for
wavelength calibration, resulting in a calibration uncertainty of $\sim$0.28~\AA. We
stacked all the frames taken with the same chip with the same mask, and extracted
spectra, and corrected for the tellulic features using the standard star spectra.

\section{Results}\label{s3}
\subsection{NIR Spectra}\label{s3-1}
With the NTT/SofI and Subaru/MOIRCS observations combined, we obtained well-exposed
spectra from a total of 23 and 44 different objects in the CBF and the Ebisawa field,
respectively. Properties of all the spectra are summarized in table~\ref{t04} and
table~\ref{t05}. The spectra are shown in figures~\ref{f05} and \ref{f06}. The
continuum was fitted with a polynomial and was subtracted, and then, normalized in an
arbitrary unit for the display.

We notice that the sources can be divided into two simply based on whether the features are
dominated by absorption or emission: (i) Sources with absorption lines such as H\emissiontype{I},
Na\emissiontype{I} (2.21 $\mu$m), Ca\emissiontype{I} (2.26 $\mu$m), and CO bandhead
features (2.29, 2.32, 2.35, 2.38 $\mu$m). (ii) Sources with emission lines such as
H\emissiontype{I} (Br$\gamma$: 2.16 $\mu$m), He\emissiontype{I} (2.11 $\mu$m), and
He\emissiontype{II} (2.19 $\mu$m),

For (i) sources, we estimated the spectral type by comparing the NIR spectral atlas of
cool stars~\citep{ali95, rayner09} assuming that they are dwarfs. We estimated a rough
spectral type for each NIR spectrum. F- and G-type stars exhibit Br$\gamma$ absorption
line (H\emissiontype{I}). G-type stars also show metallic features such as
Na\emissiontype{I}, Ca\emissiontype{I}. K- and M-type stars have molecular features of
CO and with little metallic features. The CO absorption features are dominant in the
M-type stars more than in the K-type stars. The proposed spectral types are shown in
table~\ref{t04} and \ref{t05}. For (ii), we labelled them as ``em'' in
table~\ref{t04}. The difference in the luminosity class (dwarf or giants) cannot be
distinguished with the spectral resolution in the presented study, but it does not
affect the rough spectral typing described above~\citep{rayner09}.

\begin{figure}[htbp]
 \setcounter{figure}{4}
 \begin{center}
  \FigureFile(90mm,55mm){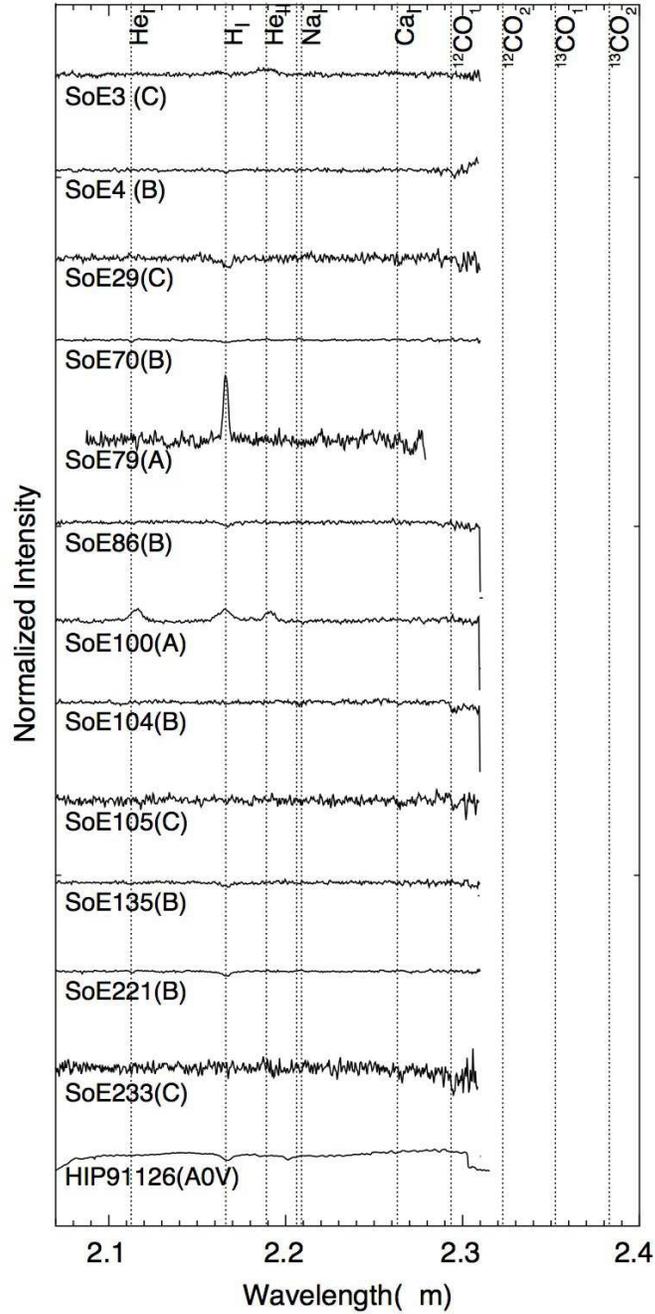}
 \end{center}
 \caption{Normalized \ksi-band spectra of sources in the Ebisawa field using NTT/SofI and
 Subaru/MOIRCS. Source numbers follow tables~\ref{t04} and \ref{t05}. SoE indicates
 that the spectra were obtained in the Ebisawa field with NTT/SofI and SuE indicates by
 Subaru/MOIRCS. The proposed class (A, B, and C; see $\S~\ref{s3-2}$) is also given for
 each spectrum. At the top, some spectroscopic features are labelled. Two $^{12}$CO and
 $^{13}$CO features are distinguished by suffix 1 and 2. At the bottom, standard star
 (A0V) spectra are shown, which are not corrected for telluric features to highlight
 these features with the \maru{+} marks.}
 \label{f05}
\end{figure}

\begin{figure}[htbp]
 \setcounter{figure}{4}
 \begin{center}
  \FigureFile(180mm,150mm){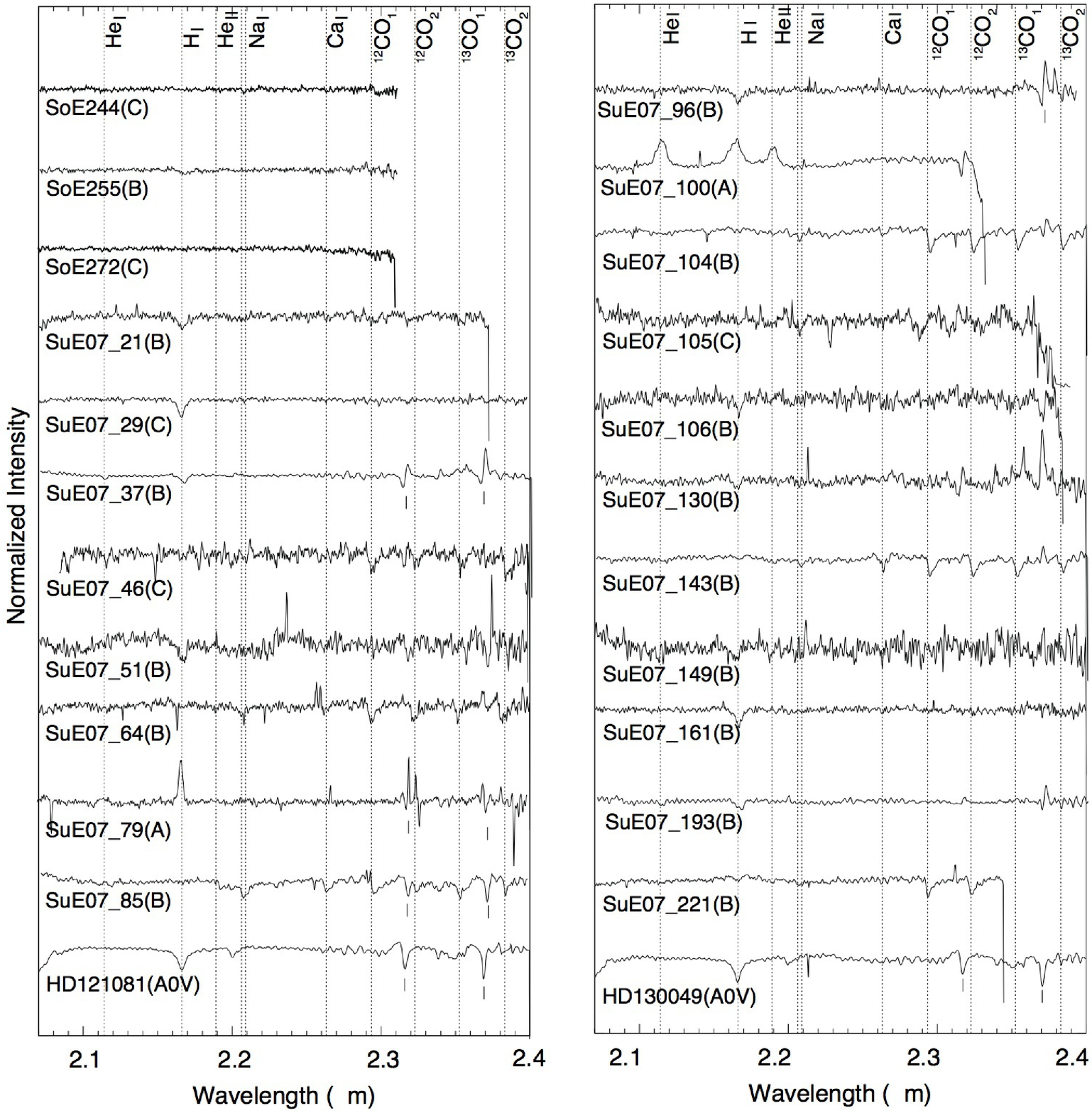}
  \caption{(Continued).}
 \end{center}
\end{figure}

\begin{figure}[htbp]
 \setcounter{figure}{4}
 \begin{center}
  \FigureFile(180mm,130mm){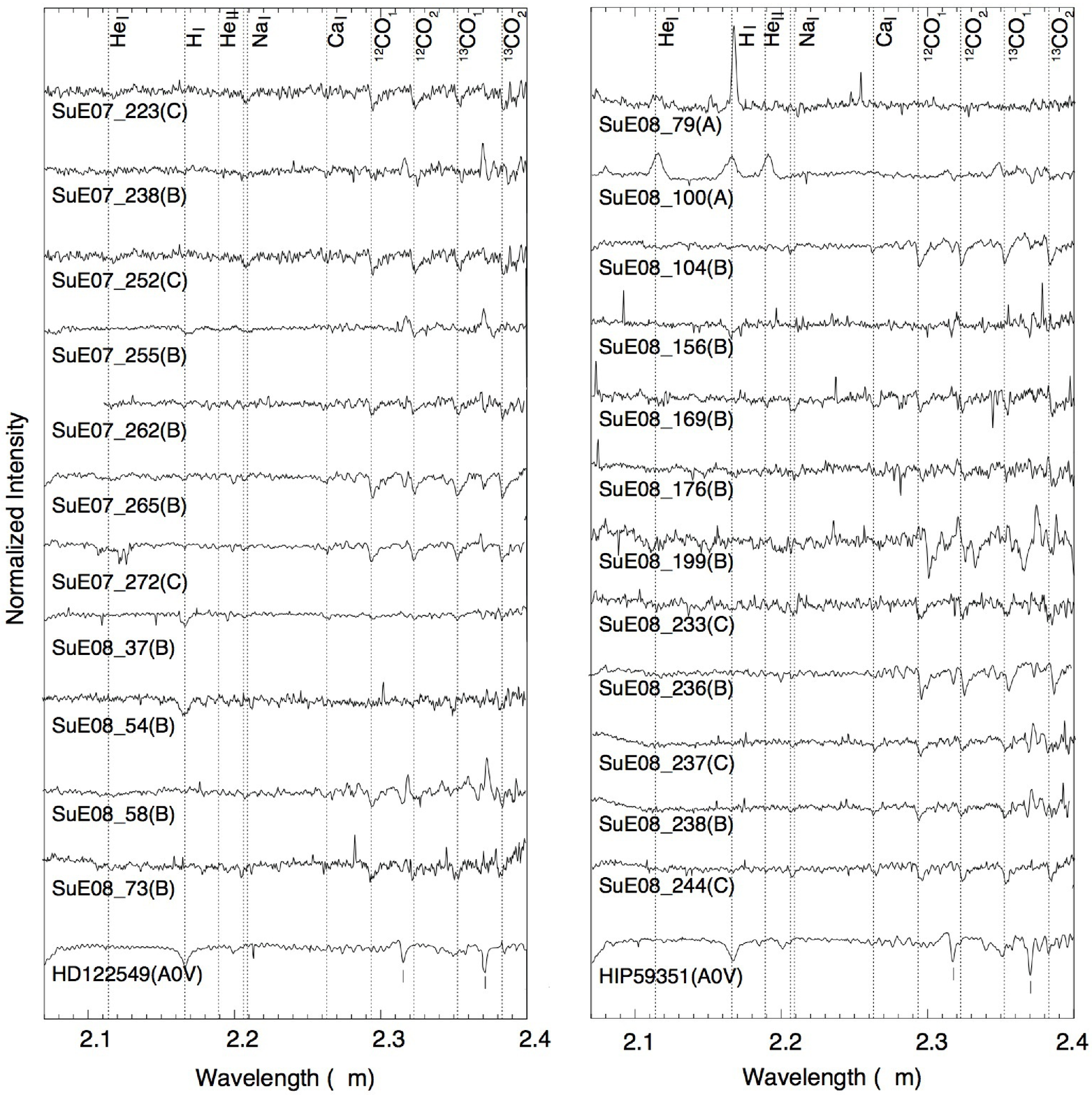}
  \caption{(Continued.)}
 \end{center}
\end{figure}

\begin{figure}[htbp]
 \setcounter{figure}{5}
 \begin{center}
  \FigureFile(180mm,100mm){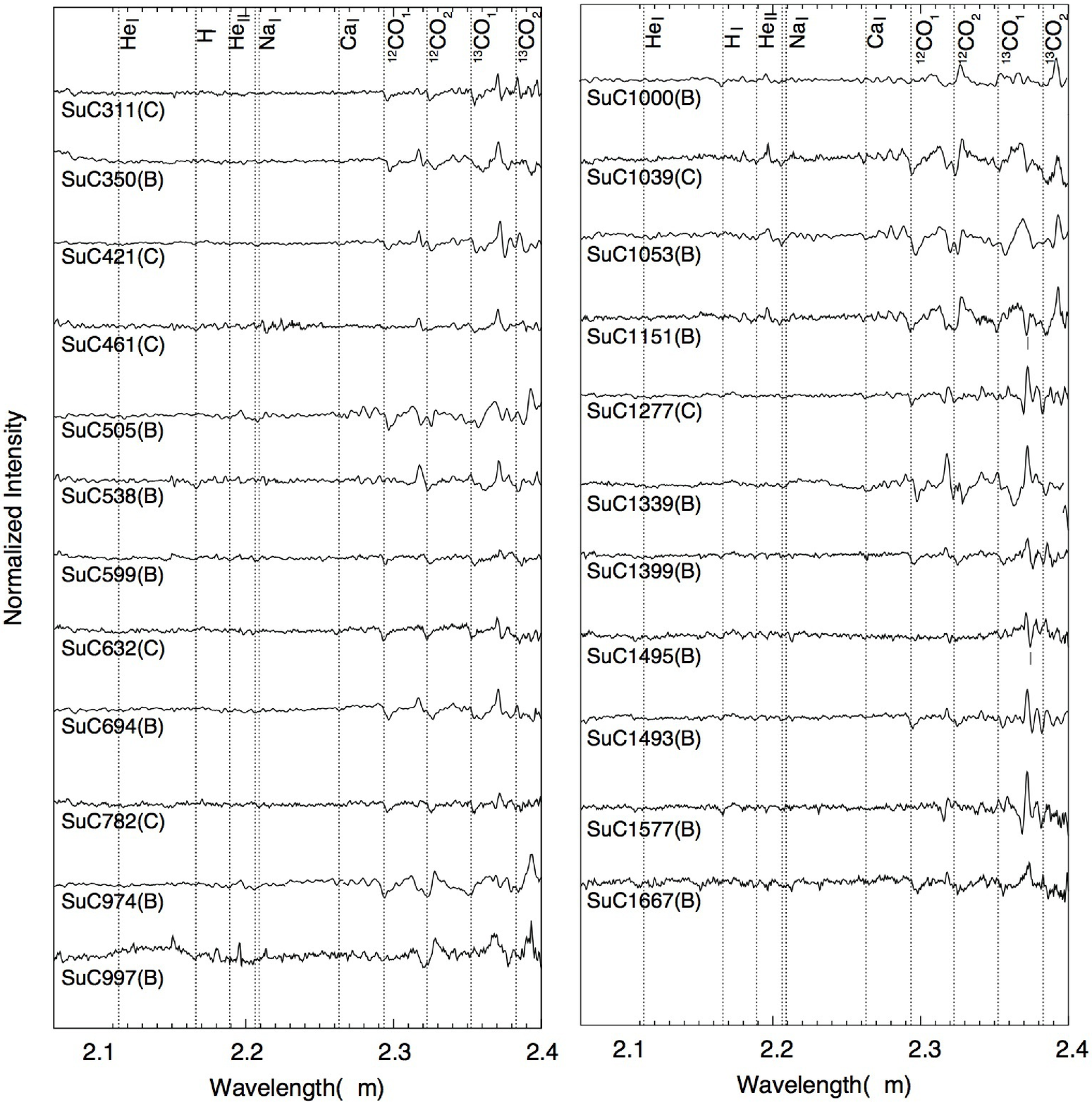}
 \end{center}
 \caption{Normalized \ksi-band spectra of sources in the CBF using Subaru/MOIRCS. 
 Source numbers follow table~\ref{t05}. Other symbols are the same with
 figure~\ref{f05}.}\label{f06}
\end{figure}

\setlength{\tabcolsep}{0.0000001in}
\begin{longtable}[bhtp,\textwidth,clip]{lcccccclccccccccclc}
  \caption{NIR spectra obtained in the Ebisawa field}\label{t04}\\
   \hline
   \hline
     Ref.\footnotemark[]{$*$}  & R.\,A. & Decl. & HR  & m$_\mathrm{\ksi}$  & \hi--\ksi \footnotemark[]{$\dagger$}  & log  & Obs\footnotemark[]{$\ddagger$} 
     & \multicolumn{9}{c}{Features\footnotemark[]{$\S$}} & Proposed\footnotemark[]{$\|$} & Class\footnotemark[]{$\#$}\\
   \cline{2-3}\cline{9-17}
   ID                              & \multicolumn{2}{c}{(J2000.0)} &    &(mag) & (mag)&  (\fx/\fks)  &  & He\emissiontype{I}\phantom{0} & H\emissiontype{I}\phantom{0} & He\emissiontype{II}\phantom{0} & Na\emissiontype{I}\phantom{0} & Ca\emissiontype{I}\phantom{0} & $^{12}$CO$_{1}$\phantom{0} & $^{12}$CO$_{2}$\phantom{0} & $^{13}$CO$_{1}$\phantom{0} & $^{13}$CO$_{2}$ & type & \\
   \hline
   \hline
   \endhead
\hline
\multicolumn{3}{l}{\hbox to 0pt{\parbox{180mm}{\footnotesize
}}}
\endfoot
\hline
\multicolumn{3}{l}{\hbox to 0pt{\parbox{180mm}{\footnotesize
      \footnotemark[$*$] Sequence ID in \citet{ebisawa05}.
      \footnotemark[$\dagger$] Non : no detection in \hi-band.
     \footnotemark[$\ddagger$] SoE : NTT/SofI observation source, SuE\_07: Subaru/MOIRCS observation source, SuE\_08:Subaru/MOIRCS 2008 observation source.
     \footnotemark[$\S$] + shows absorption features and -- shows emission features.
     \footnotemark[$\|$] Proposed spectral types based on each spectral features. ? : difficult to determine spectral type with low S/N features. em : spectra with emission features such as H\emissiontype{I}, He\emissiontype{I}, and He\emissiontype{II}.
     \footnotemark[$\#$] A, B, and C indicate the proposed class we defined in $\S~\ref{s3-2}$.
}}}
\endlastfoot
3    & 18:42:58.3 & $-$03:53:27 &  \phantom{--}\phantom{--}\phantom{-}0.86 & 12.5 & 2.20 & $-$1.65 & SoE & \phantom{--}& + & \phantom{--} & \phantom{--} & \phantom{--} & \phantom{--} &\phantom{--}  & \phantom{--} & \phantom{--} & late G &C\\
4    & 18:43:00.4 & $-$03:53:49 &  \phantom{--}$-$0.03 & 12.1 & 1.60 & $-$2.24 & SoE          & \phantom{--} & \phantom{--}& \phantom{--} & \phantom{--} & \phantom{--} & +  & \phantom{--}&\phantom{--}  & \phantom{--}& early K&B\\
21   & 18:43:13.8 & $-$03:57:07 &  \phantom{--}$-$0.24 & 12.7 & 0.34 & $-$3.14& SuE\_07  &\phantom{--}  &  +          & \phantom{--} & \phantom{--} & \phantom{--} & +  & \phantom{--}&\phantom{--}  & \phantom{--} & early K &B\\
29   & 18:43:17.5 & $-$03:56:00 &  \phantom{--}\phantom{--}\phantom{-}0.29 & 12.0 & 0.14 & $-$2.74 & SoE         & \phantom{--} & + &  \phantom{--}& \phantom{--}& \phantom{--} & \phantom{--} & \phantom{--} & \phantom{--} & \phantom{--} & early K&C\\
     &            &             &         &      &      &          & SuE\_07     & \phantom{--} & + &  \phantom{--}& \phantom{--}& \phantom{--} & \phantom{--} & \phantom{--} & \phantom{--} & \phantom{--} & early K &C \\
37 & 18:43:19.0 & $-$03:53:27 &  \phantom{--}$-$0.54 & 12.8 &0.18  &$-$3.69 & SuE\_07 & \phantom{--} & +  & \phantom{--} &  \phantom{--}  & \phantom{--}  & \phantom{--}  & \phantom{--}  & \phantom{--} & \phantom{--} & late K&B\\
     &            &             &         &      &      &          & SuE\_08     & \phantom{--} & + &  \phantom{--}& \phantom{--}& \phantom{--} & + & + & \phantom{--} & \phantom{--} & late K &B \\
46   & 18:43:21.1 & $-$03:54:30  &  \phantom{--}\phantom{--}\phantom{-}0.60 & 14.6  &  -- & $-$1.60 & SuE\_07  & \phantom{--} & + &  \phantom{--}& \phantom{--}& \phantom{--} & \phantom{--} & \phantom{--} & \phantom{--} & \phantom{--} & early K &C\\
51   & 18:43:21.8 & $-$03:53:03 &  \phantom{--}$-$0.40 & 13.3 & 0.23 & $-$2.44 & SuE\_07 &\phantom{--} & + &  \phantom{--}& \phantom{--}& \phantom{--} & \phantom{--} & \phantom{--} & \phantom{--} & \phantom{--}&  early K&B\\
54   & 18:43:23.0 & $-$03:57:53 &  \phantom{--}$-$0.44 & 13.3 & 0.23 & $-$2.93 & SuE\_08 & \phantom{--} & + &  \phantom{--}& \phantom{--}& \phantom{--} & \phantom{--} & \phantom{--} & \phantom{--} & \phantom{--}& early K&B\\
58   & 18:43:23.6 & $-$03:53:14  & \phantom{--}\phantom{-}0.09 & 14.4 & 0.19 & $-$2.27 & SuE\_08 & \phantom{--} & \phantom{--} &  \phantom{--}& \phantom{--}& \phantom{--} & + & \phantom{--} & \phantom{--} & \phantom{--}& late K--early M &B\\
64   & 18:43:24.5 & $-$03:53:50 &  \phantom{--}$-$0.31 & 12.2 & 0.43 & $-$2.21 & SuE\_07 & \phantom{--} & \phantom{--} &  \phantom{--}& \phantom{--}& & + & + & + & + &late K--early M &B\\
70   & 18:43:28.4 & $-$04:07:33 &  \phantom{--}$-$0.72 &  8.5 &0.15  & $-$3.94 & SoE         & + & + & \phantom{--} & \phantom{--} & \phantom{--} & \phantom{--} & \phantom{--} & \phantom{--} &\phantom{--}  & late G&B \\
73   & 18:43:28.9 & $-$03:57:33 &  \phantom{--}$-$0.56 & 13.2 & 0.34 & $-$2.47 & SuE\_08 & \phantom{--} & + &  \phantom{--}& \phantom{--}& \phantom{--} & + & + & + & \phantom{--}& late K--early M &B\\
79   & 18:43:29.7 & $-$03:50:15 &  \phantom{--}\phantom{--}\phantom{-}0.99 & 12.9 & Non & $-$2.07 & SoE& \phantom{--} & -- & \phantom{--} & \phantom{--} &\phantom{--}  & \phantom{--}& \phantom{--} & \phantom{--} & \phantom{--} &em&A\\
    &  &    &   &  &  &  & SuE\_07& \phantom{--} & -- & \phantom{--} & \phantom{--} &\phantom{--}  & \phantom{--}& \phantom{--} & \phantom{--} & \phantom{--} &em &A\\
    &  &  &  &  &  & & SuE\_08& \phantom{--} & -- & \phantom{--} & \phantom{--} &\phantom{--}  & \phantom{--}& \phantom{--} & \phantom{--} & \phantom{--} & em&A\\
85   & 18:43:30.6 & $-$03:53:52 &  \phantom{--}$-$1.00 & 11.0 & 0.26 & $-$4.13 & SuE\_07 & \phantom{--} & + &  \phantom{--}& + & + & + & \phantom{--} & \phantom{--} & \phantom{--}& late K&B\\
86   & 18:43:30.8 & $-$04:01:03 &  \phantom{--}$-$0.74 &  9.3 &0.13  & $-$3.97  & SoE         & \phantom{--} & + & \phantom{--} & \phantom{--} & \phantom{--} & \phantom{--} & \phantom{--} & \phantom{--} & \phantom{--} &late G&B\\
96   & 18:43:31.8 & $-$03:57:17  &  \phantom{--}$-$0.33 &  13.9 & 1.70  & $-$2.72 & SuE\_07 & \phantom{--} & + & \phantom{--} & \phantom{--} & \phantom{--} & \phantom{--} & \phantom{--} & \phantom{--} & \phantom{--} & early K &B\\
100  & 18:43:32.6 & $-$04:04:19 &  \phantom{--}\phantom{--}\phantom{-}0.74 & 11.1 & 1.41 & $-$1.97 & SoE & -- & -- & -- & \phantom{--} & \phantom{--} &\phantom{--}  & \phantom{--} & \phantom{--} & \phantom{--} & em &A\\
     & &  &     &  &  &  & SuE\_07 & -- & -- & --  &  &  &   &  &   &   & em &A\\
     & &  &     &  &  &  & SuE\_08 & -- & --  & --  &  &  &   &  &   &   & em &A\\
104  & 18:43:33.5 & $-$04:03:54 &  \phantom{--}$-$0.29 & 10.4 & 0.36 & $-$3.44  & SoE     & \phantom{--}&  \phantom{--}& \phantom{--} & \phantom{--} &  \phantom{--}     & +  & \phantom{--} & \phantom{--} & \phantom{--} &early M&B\\
     & &  &     &  &  &  & SuE\_07 &  &  &   &  &  & +  & + & +  & +  & early M&B\\
     & &  &     &  &  &  & SuE\_08 &  &  &   &  &  & +  & + & +  & +  & early M&B\\
105  & 18:43:33.9 & $-$03:52:53 &  \phantom{--}\phantom{--}\phantom{-}0.70 & 13.1 &1.80  & $-$2.42 & SoE &\phantom{--} & + & \phantom{--} & \phantom{--} & \phantom{--} & \phantom{--} &  \phantom{--}& \phantom{--} & \phantom{--} & ? &C\\
  &   &  &  &   &   &  & SuE\_07 &\phantom{--} & \phantom{--} & \phantom{--} & + & \phantom{--} & \phantom{--} &  \phantom{--}& \phantom{--} & \phantom{--} & late K--early M &C\\
106 & 18:43:34.1 & $-$03:55:24 &  \phantom{--}$-$0.67 & 14.1 & 0.37 & $-$2.90 & SuE\_07 &  \phantom{--} & + & \phantom{--} & \phantom{--} & \phantom{--} & \phantom{--} & \phantom{--} & \phantom{--} & \phantom{--} & early K &B\\
130 & 18:43:49.5 & $-$03:52:33 &   \phantom{--}$-$0.60 & 12.6 & 0.30 & $-$3.24  & SuE\_07 & \phantom{--} & + & \phantom{--} & \phantom{--} &\phantom{--} & \phantom{--} &\phantom{--}  & \phantom{--} & \phantom{--} & late G--early K &B\\
135  & 18:43:39.2 & $-$03:52:53 &  \phantom{--}$-$0.79 &  9.9 & 0.36 & $-$4.05 & SoE        & \phantom{--} & + & \phantom{--} & \phantom{--} & \phantom{--}& \phantom{--} & \phantom{--} & \phantom{--} & \phantom{--} &early K &B\\
143  & 18:43:41.0   & $-$03:58:02 &  \phantom{--}$-$0.50 & 10.7 & 0.45 & $-$3.70 & SuE\_07 & \phantom{--} & + & \phantom{--} & + & +& + & + & + & + &late K--early M &B\\
149  & 18:43:42.6 & $-$03:59:41 &  \phantom{--}$-$0.13 & 14.5 & 0.19 & $-$1.97 & SuE\_07 & \phantom{--} & + & \phantom{--} & \phantom{--} &\phantom{--} & \phantom{--} &\phantom{--}  & \phantom{--} & \phantom{--} & late G--early K &B\\
156  & 18:43:45.4 & $-$03:53:17 &  \phantom{--}$-$0.60 & 13.8 & 0.27 & $-$2.26& SuE\_08 & \phantom{--} & + & \phantom{--} & \phantom{--} &\phantom{--} & \phantom{--} &\phantom{--}  & \phantom{--} & \phantom{--} & late G--early K&B\\
161 & 18:43:46.4 & $-$03:54:12 &  \phantom{--}$-$0.83 & 13.4 & 0.17 & $-$2.95 & SuE\_07 & \phantom{--} & + & \phantom{--} & \phantom{--} &\phantom{--} & \phantom{--} &\phantom{--}  & \phantom{--} & \phantom{--} & late G--early K &B\\
169 & 18:43:47.1 & $-$03:53:18 &  \phantom{--}$-$0.29 & 14.2 &0.37 & $-$1.88 & SuE\_08 & \phantom{--} & \phantom{--}  & \phantom{--} &  + & \phantom{--}  & + & +  & \phantom{--} & \phantom{--} &late G- early K &B\\
176 & 18:43:48.7 & $-$04:01:36 &  \phantom{--}$-$0.56 & 13.4 & 0.24  & $-$2.36 & SuE\_08 &\phantom{--} & +  & \phantom{--} &  + & +  & \phantom{--}  & \phantom{--}  & \phantom{--} & \phantom{--} & K &B\\
193 & 18:43:53.1 & $-$03:57:60 &  \phantom{--}$-$0.16 & 12.7 & 0.22 & $-$2.66 & SuE\_07 & \phantom{--} & + & \phantom{--} & \phantom{--} &\phantom{--} & \phantom{--} &\phantom{--}  & \phantom{--} & \phantom{--} & late G--early K &B\\
199 & 18:43:54.8 & $-$04:07:42 &  \phantom{--}\phantom{--}\phantom{-}0.06 & 13.3 & 0.25 & $-$2.36 & SuE\_08 & \phantom{--} & + & \phantom{--} & \phantom{--} &\phantom{--} & + & +  & + & +& late K--M &B\\
221  & 18:43:59.7 & $-$03:55:18 &  \phantom{--}$-$0.53 &  8.8 & 0.17 & $-$4.20  & SoE        & \phantom{--}& + & \phantom{--} & \phantom{--} &\phantom{--}  & \phantom{--} & \phantom{--} & \phantom{--} & \phantom{--} &early K &B\\
       &    &   &  &   &  &   & SuE\_07                                            & \phantom{--}& + & \phantom{--} & \phantom{--} &\phantom{--}  & + & + & \phantom{--} & \phantom{--} &early K--late K &B\\
223 & 18:44:00.3 & $-$04:05:59 &  \phantom{--}\phantom{--}\phantom{-}0.50 & 11.9 &0.46  &$-$2.91 & SuE\_07 & \phantom{--}& + & \phantom{--} & + &\phantom{--}  & + & + & + & +& late K--early M &C\\
233  & 18:44:03.9 & $-$04:02:58 &  \phantom{--}\phantom{--}\phantom{-}0.25 & 13.4 & 0.82 & $-$1.46  &SoE & \phantom{--}& \phantom{--} &  \phantom{--}&  \phantom{--}&  \phantom{--}& +&\phantom{--} &\phantom{--}  & \phantom{--} & lateK--early M &C\\
        &         &  &  &  &  &   &SuE\_07                                          & \phantom{--}& \phantom{--} &  \phantom{--}&  + &  \phantom{--}& + & +  & +  & + & late K--early M &C\\
       &         &  &  &  &  &   &SuE\_08 & \phantom{--}& \phantom{--} &  \phantom{--}&  \phantom{--} &  \phantom{--}& + & + & + & + & late K--early M &C\\
236  & 18:44:05.4 & $-$04:08:32 &  \phantom{--}$-$0.73 & 14.1 & 0.24 & $-$2.05& SuE\_08 & \phantom{--}& \phantom{--} &  \phantom{--}& \phantom{--}  &  \phantom{--}& + & + & + & +  & late K -early M &B\\
237 & 18:44:05.6 & $-$04:05:39 &  \phantom{--}\phantom{--}\phantom{-}0.14 & 14.9 & 0.41 & $-$1.81 & SuE\_08 &\phantom{--}& + &  \phantom{--}& +  &  + & + & + & + & \phantom{--} &  late K--early M  &C\\
238 & 18:44:05.9 & $-$04:06:12 &  \phantom{--}\phantom{--}\phantom{-}0.01 & 13.1 & 0.26 & $-$2.12 & SuE\_07 & \phantom{--}& + &  \phantom{--}& +  &  + & + & + & + & \phantom{--} & late K--early M &B\\
       &  &  &  &  &  &  & SuE\_08 & \phantom{--}& + &  \phantom{--}& +  &  + & + & + & + & + & late K--early M &B\\
244  & 18:44:11.0 & $-$04:05:17 & \phantom{--}\phantom{-}0.48 & 12.7 & 0.84 & $-$2.20  &SoE & \phantom{--}&\phantom{--}  & \phantom{--} & + &\phantom{--}  & \phantom{--} & \phantom{--} & \phantom{--} & \phantom{--}& late K--early M &C\\
     &         &  &  &  &  &   &SuE\_08     & \phantom{--}& \phantom{--} &  \phantom{--}&  + &  \phantom{--}& + & +  & +  & + & late K--early M &C\\
252 & 18:44:15.6 & $-$04:03:57 & \phantom{--}\phantom{-}0.79 & 13.6 & 0.99 & $-$1.98 & SuE\_07 &\phantom{--}& +  & \phantom{--} & + &\phantom{--}  & + & + & + & \phantom{--}& late K--early M &C\\                                                     &  &  &  &  &  &   &SuE\_08 & \phantom{--}&\phantom{--}  & \phantom{--} & \phantom{--} &\phantom{--}  & + & + & + & + & late K-- M &C\\
255  & 18:44:18.6 & $-$04:06:03 &  \phantom{--}\phantom{--}\phantom{-}0.06 & 12.1 & 0.48 & $-$2.60  &SoE & \phantom{--}& +  & \phantom{--} & \phantom{--} &\phantom{--}  & \phantom{--} & \phantom{--} & \phantom{--} & \phantom{--}& K &B\\
     &  &  &  &  &  &   &SuE\_07 & \phantom{--}& +  & \phantom{--} & + &\phantom{--}  & \phantom{--} & \phantom{--} & \phantom{--} & \phantom{--}& K &B\\
262  & 18:44:22.7 & $-$04:00:17 & \phantom{--}\phantom{-}0.10 & 12.5 & 0.87 & $-$2.56 & SuE\_07 &  \phantom{--}& \phantom{--}  & \phantom{--} & \phantom{--}&\phantom{--}  & + & + & + & \phantom{--}&late M &B\\
265  & 18:44:24.2 & $-$04:01:35 &  \phantom{--}$-$0.09 & 13.8 & 0.87 & $-$2.03 & SuE\_07 & \phantom{--}& \phantom{--}  & \phantom{--} & \phantom{--}&\phantom{--}  & + & + & + & + & late M&B\\
272  & 18:44:28.8 & $-$04:01:03 &  \phantom{--}\phantom{--}\phantom{-}0.27          & 11.1 & 0.72 & $-$2.37 &SoE & \phantom{--}&\phantom{--}  & \phantom{--} & \phantom{--} &\phantom{--}  & + & \phantom{--} & \phantom{--} & \phantom{--} & late K--M &C\\
     &     &  &    &  &  &   & SuE\_07 & \phantom{--}&\phantom{--}  & \phantom{--} & \phantom{--} &\phantom{--}  & + & + & + & + & late K--M &C\\
\hline
\end{longtable}

\setlength{\tabcolsep}{0.00000001in}
\begin{longtable}[bhtp,\textwidth,clip]{lccccccccccccccclc}
  \caption{NIR Spectra obtained in the Chandra Bulge Field}\label{t05}\\
   \hline
   \hline
     Ref.\footnotemark[]{$*$}  & R.\,A. & Decl. & HR   & m$_\mathrm{\ksi}$  & \hi--\ksi\footnotemark[]{$\dagger$} & log & \multicolumn{9}{c}{Features\footnotemark[]{$\ddagger$}} & Proposed\footnotemark[]{$\S$} & Class\footnotemark[]{$\|$}\\
   \cline{2-3}\cline{8-16}
   ID                              & \multicolumn{2}{c}{(J2000.0)} &    & (mag) & (mag) & (\fx/\fks) & He\emissiontype{I}\phantom{0} & H\emissiontype{I}\phantom{0} 
   & He\emissiontype{II}\phantom{0} & Na\emissiontype{I}\phantom{0} & Ca\emissiontype{I}\phantom{0} & $^{12}$CO$_{1}$\phantom{0} & $^{12}$CO$_{2}$\phantom{0} & 
   $^{13}$CO$_{1}$\phantom{0} & $^{13}$CO$_{2}$ & type & \\
   \hline
   \hline
   \endhead
\hline
\endfoot
\hline
\multicolumn{3}{l}{\hbox to 0pt{\parbox{170mm}{\footnotesize
      \footnotemark[$*$] Sequence ID in \citet{morihana13}.
      \footnotemark[$\dagger$] Non : no detection in \hi-band.
     \footnotemark[$\ddagger$] + shows absorption features and -- shows emission features.
    \footnotemark[$\S$] Proposed spectral types based on each spectral features. ? : difficult to determine spectral type. 
    \footnotemark[$\|$] Class A, B, and C indicate proposed class we defined in $\S~\ref{s3-2}$.
}}}
\endlastfoot
311 & 17:51:09.2 & $-$29:39:39 &   \phantom{--}\phantom{--}\phantom{-}0.42 & 11.7 & 0.29  &$-$3.34  & \phantom{--} & \phantom{--}& \phantom{--} & \phantom{--} & \phantom{--} & +  & + & + & \phantom{--}& late K  & C\\
350 & 17:51:10.7 & $-$29:41:40 &  \phantom{--}$-$0.17 & 11.8 & 0.32 &$-$3.22  & \phantom{--} & \phantom{--}& \phantom{--} & \phantom{--} & \phantom{--} & +  & + & \phantom{--} & \phantom{--}& early M  & B\\
421 & 17:51:13.3 & $-$29:29:31 &  \phantom{--}\phantom{--}\phantom{-}0.41 & 13.9 & 0.67  &$-$2.41  & \phantom{--} & + & \phantom{--} & + & + & +  & + & + & \phantom{--}   & late K--early M & C\\
461 & 17:51:14.8 & $-$29:38:50 &  \phantom{--}\phantom{--}\phantom{-}0.38 & 13.0 & 0.29  &$-$2.84  & \phantom{--} & + & \phantom{--} & \phantom{--} & \phantom{--} & \phantom{--}  & \phantom{--} & \phantom{--} & \phantom{--}  & early K & C\\
505 & 17:51:16.3 & $-$29:34:46 &  \phantom{--}\phantom{--}\phantom{-}0.10 & 14.0 & 0.28 & $-$2.95 &\phantom{--} & \phantom{--}& \phantom{--} & + & \phantom{--} & \phantom{--}  & \phantom{--} & \phantom{--} & \phantom{--}  & K  & B\\
538 & 17:51:17.2 & $-$29:38:23 &  \phantom{--}$-$0.95 & 12.0 & 0.06  & $-$4.13 & \phantom{--} & + & \phantom{--} & \phantom{--} & \phantom{--} & \phantom{--}  & + & \phantom{--} & \phantom{--} &  K  & B\\
599 & 17:51:18.6 & $-$29:30:15 &  \phantom{--}$-$0.26 & 11.1 &0.37  & $-$3.84  & \phantom{--} & \phantom{--} & \phantom{--} & \phantom{--} & \phantom{--} & +  & + & + & \phantom{--} & late K--early M & B\\
632 & 17:51:19.4 & $-$29:42:46 &  \phantom{--}\phantom{--}\phantom{-}0.91 & 13.3 & 0.41 & $-$2.54  & \phantom{--} & \phantom{--} & \phantom{--} & \phantom{--} & \phantom{--} & +  & + & + & \phantom{--}   & late K--early M & C\\
694 & 17:51:20.7 & $-$29:28:56 &  \phantom{--}$-$0.09 & 10.8 &0.50 &$-$3.95  &\phantom{--} & \phantom{--} & \phantom{--} & \phantom{--} & \phantom{--} & +  & + & + & \phantom{--} & late K--M  & B\\
782 & 17:51:22.6 & $-$29:42:54 &  \phantom{--}\phantom{--}\phantom{-}0.52 & 13.3 &0.24 & $-$2.16   &\phantom{--} & \phantom{--} & \phantom{--} & \phantom{--} & \phantom{--} & +  & + & + & \phantom{--}&  late K--M & C\\
974 & 17:51:27.5 & $-$29:33:51 & \phantom{--}$-$0.76 & 11.4 & 0.39 & $-$3.82  & \phantom{--} & \phantom{--} & \phantom{--} & \phantom{--} & \phantom{--} & +  & + & + & + & M  & B\\
997 & 17:51:46.5 & $-$29:34:01 &  \phantom{--}$-$0.97 & 13.5 & Non & $-$3.35  & \phantom{--} & \phantom{--} & \phantom{--} & \phantom{--} & \phantom{--} &  \phantom{--} & + & \phantom{--} &  \phantom{--}& M  & B\\
1000 & 17:51:28.1 & $-$29:37:03 &  \phantom{--}$-$0.84 & 10.9 & 0.03 &$-$4.31  & \phantom{--} & + & \phantom{--} & \phantom{--} & \phantom{--} & \phantom{--}  &\phantom{--}  & \phantom{--} & \phantom{--} &  early K  & B\\
1039 & 17:51:29.1 & $-$29:34:44 &  \phantom{--}\phantom{--}\phantom{-}0.20 & 12.5  & 0.38 & $-$3.46  & \phantom{--} &\phantom{--}  & \phantom{--} & + & \phantom{--} & +  & +  & + & \phantom{--} &late K  & C\\
1053 & 17:51:29.4 & $-$29:37:46 &  \phantom{--}\phantom{--}\phantom{-}0.01 & 10.8  & 0.43 & $-$4.57  & \phantom{--} &\phantom{--}  & \phantom{--} & + & \phantom{--} & +  & +  & \phantom{--} & \phantom{--}  & M  & B\\
1151 & 17:51:31.8 & $-$29:33:51 &  \phantom{--}$-$0.12 & 12.7 & 0.29  & $-$3.46  & \phantom{--} &\phantom{--}  & \phantom{--} & + & \phantom{--} & +  & +  & + & \phantom{--} & M  & B\\
1277 & 17:51:34.5 & $-$29:36:09 &  \phantom{--}\phantom{--}\phantom{-}0.24 & 13.8  & 0.14 & $-$2.60  & \phantom{--} &\phantom{--}  & \phantom{--} & + & \phantom{--} & +  & \phantom{--}  & \phantom{--} & \phantom{--}& M & C\\
1339 & 17:51:36.4 & $-$29:32:17 &  \phantom{--}$-$0.05 & 13.1 & 0.47 & $-$2.96  & \phantom{--} &\phantom{--}  & \phantom{--} & + & \phantom{--} & +  & +  & \phantom{--} & \phantom{--}&  M & B\\
1399 & 17:51:37.7 & $-$29:33:05 &  \phantom{--}$-$0.78 & 11.5 & 0.24 & $-$3.92  & \phantom{--} &\phantom{--}  & \phantom{--} & + & \phantom{--} & +  & +  & + & \phantom{--}& early M & B\\
1495 & 17:51:39.8 & $-$29:37:10 &  \phantom{--}$-$0.82 & 13.3 & 0.02  & $-$3.06  & \phantom{--} &\phantom{--}  & \phantom{--} & + & \phantom{--} & \phantom{--} & \phantom{--}  & \phantom{--} & \phantom{--}& ? & B\\
1493 & 17:51:39.8 & $-$29:33:18 &  \phantom{--}$-$0.19 & 11.7 & 0.26 & $-$3.96  & \phantom{--} &\phantom{--}  & \phantom{--} & + & \phantom{--} & + & +  & \phantom{--} & \phantom{--}&early K & B\\
1577 & 17:51:42.1 & $-$29:35:37 &  \phantom{--}\phantom{--}\phantom{-}0.06 & 13.1 & 0.06 &$-$3.63   &  \phantom{--} & +  & \phantom{--} &\phantom{--}  & \phantom{--} & \phantom{--} & \phantom{--}  & \phantom{--} & \phantom{--}&early K  & B\\
1667 & 17:51:44.8 & $-$29:37:12 &  \phantom{--}$-$0.15 & 13.0 & 0.32 & $-$3.86  &\phantom{--} & \phantom{--}  & \phantom{--} &\phantom{--}  & \phantom{--} & + & + & + & \phantom{--}& early M & B\\
\hline
\end{longtable}

\subsection{X-ray Spectra}\label{s3-2}
Since most X-ray sources are too dim to make individual X-ray spectrum, we use hardness
ratio to characterize X-ray spectra. We define the spectral hardness ratio as HR =
(H-S)/(H+S), where H and S are the normalized count rate in the hard (2--8 keV) and soft
(0.5--2 keV) energy band.  We calculated the HR for each source in the both fields and
show them in tables~\ref{t04} and \ref{t05}. Based on X-ray hardness and NIR spectral
features, we can classify sources into three: those with (A) hard X-ray spectra
(HR$\geq$0.11) and NIR emission features such as H\emissiontype{I}
(Br$\gamma$), He\emissiontype{I}, and He\emissiontype{II} (2 objects), (B) soft X-ray
spectra (HR$<$0.11) and NIR absorption features such as
H\emissiontype{I}, Na\emissiontype{I}, Ca\emissiontype{I} and CO bandheads (46 objects),
and (C) hard X-ray spectra (HR$\geq$0.11) and NIR absorption features
such as H\emissiontype{I}, Na\emissiontype{I}, Ca\emissiontype{I} and CO bandheads (17
objects).

\begin{figure}[htbp]
 \begin{center}
  \includegraphics[width=160mm]{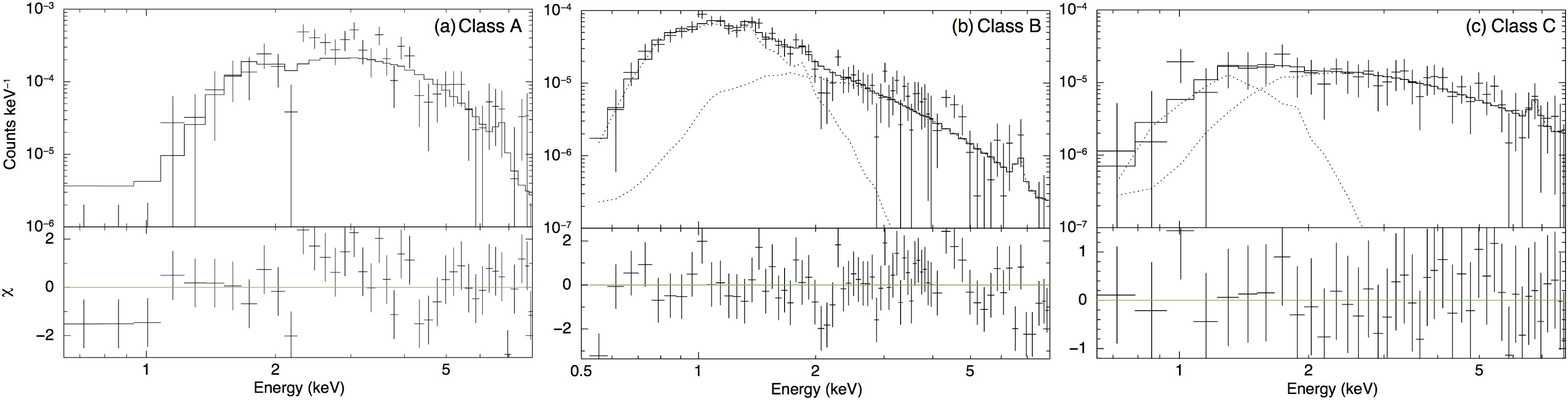}
 \end{center}
 \caption{Composite spectra and the best-fit models of the three classes. Grouped data
 (pluses) and the best-fit model convolved with the instrumental response (solid
 histograms) are shown in the upper panel, while the residuals to the fit are shown in
 the lower panel. The best-fit parameters are given in table~\ref{t06}.}\label{f07}
\end{figure}

To characterize the X-ray spectral features further, we made a
composite X-ray spectrum in the 0.5--8.0~keV of each class (figure~\ref{f07}). Note that
the composite spectrum is the sum (not the average) of the spectra of individual sources. 
Each X-ray spectra in 0.5- 8 keV were binned so that each bin includes more than  20 counts 
in order to use the $\chi$$^{2}$ statistics. 
For the spectral models, we first used a one-temperature optically-thin thermal plasma model
(\rm{APEC}, \cite{smith01}) convolved with an interstellar absorption model \citep{Wilms00} with the
interstellar abundance by \citet{anders89}. This thermal model includes the Fe K line, of which strength is characterized with  
metal abundance parameter relative to solar abundance (Z) (table~\ref{t06}). Though the fits are barely  acceptable, they give a general 
idea of the spectral shape, such that the composite spectrum of class A source seems to be best described by a single high 
temperature component, while class B and C seem to require an additional component. We then added another one-temperature 
plasma model. From the fitting, it is found that the X-ray spectra of class A and C are dominated by the higher 
temperature component, while that of the class B is by the lower temperature component.

\begin{table}[htbp]
 \caption{Best-fit parameters for global spectral fittings\footnotemark[]{$*$}}\label{t06}
\begin{center}
 \begin{tabular}{llllll}
  \hline
   \hline
   Class  &  ~\nh                                             & \kt$_{1}$  & \kt$_{2}$ & Z\footnotemark[]{$\dagger$} & $\chi^{2}$/d.o.f \\
              &   ($\times$10$^{22}$~cm$^{-2}$) & (keV)       & (keV)       &    &   \\
\hline
A   &  3.66$^{+0.74}_{-0.53}$  & 3.63$_{-1.12}^{+1.35}$  &  --    & 0.21$_{-0.15}^{+0.78}$ & \phantom{0}57.54/38\\
B   & 1.38$_{-0.10}^{+0.19}$  & 0.27$_{-0.08}^{+0.08}$   & 2.04$_{-0.53}^{+0.74}$   & 0.15$_{-0.10}^{+0.32}$   & \phantom{0}82.21/64\\
C   & 2.32$_{-0.78}^{+0.87}$  & 0.21$_{-0.09}^{+0.20}$   & 5.37$_{-2.17}^{+1.95}$   &  0.24$_{-0.20}^{+0.49}$  & \phantom{0}13.29/35\\
\hline
\multicolumn{2}{@{}l@{}}{\hbox to 0pt{\parbox{135mm}{
   \footnotesize
   \par \noindent
   \footnotemark[$*$]The best-fit value and a 1$\sigma$ statistical uncertainty are given.\\
   \footnotemark[$\dagger$] Metal abundance relative to the solar value.\\
   }\hss}}
\end{tabular}
\end{center}
\end{table}

\section{Discussion}\label{s4}
\subsection{Classification}\label{s4-1}
We discuss the nature of the sources for the three classes defined in
\S~\ref{s3-2}. First, we argue that class A sources (ID 79 and 100 in Ebisawa field) are
CVs for the following reasons : (i) They are hard X-ray sources with large HR values (HR of ID 79
and 100 are 0.71 and 0.11, respectively). (ii) A probable Fe K feature and a dominant
high-temperature ($\sim$4 keV) plasma (table~\ref{t06}) are seen in the composite X-ray
spectrum. (iii) They also exhibit NIR spectra with H\emissiontype{I} (Br$\gamma$),
He\emissiontype{I}, and He\emissiontype{II} emission lines, which presumably arise from
the accretion disk. All of these are the established observational characteristics of CVs (e.g.,
\cite{ezuka99}, \cite{baskil05}, \cite{dhillon97}).

Second, we argue that class B sources are late-type stars with enhanced coronal X-ray 
activities (X-ray active stars) for the following reasons : (i) They show a soft X-ray spectrum (table~\ref{t06}),
in which the low-temperature (0.3~keV) component is dominant. The low-temperature
component is presumably from the stellar corona, and the high-temperature component is
from flares. (ii) They exhibit NIR spectra with the absorption lines commonly seen from the
cool stars such as H\emissiontype{I}, Na\emissiontype{I}, Ca\emissiontype{I}, and CO bandheads.

The presence of the class A and B populations is no surprise, which is perfectly
compatible with current understandings of the population of faint Galactic X-ray
point sources as discussed in $\S$\ref{s1}. Before starting the presented study, we had
anticipated that most of our sources would be from these classes. Now, class C appeared
as an unexpected new class, which has not been recognized so far. Rather surprisingly,
only 2 of the 19 NIR spectra of hard X-ray sources are class A (CV) whereas a majority
of them (17 of 19 sources) are class C.

\begin{figure}[htbp]
 \begin{center}
  \includegraphics[width=150mm]{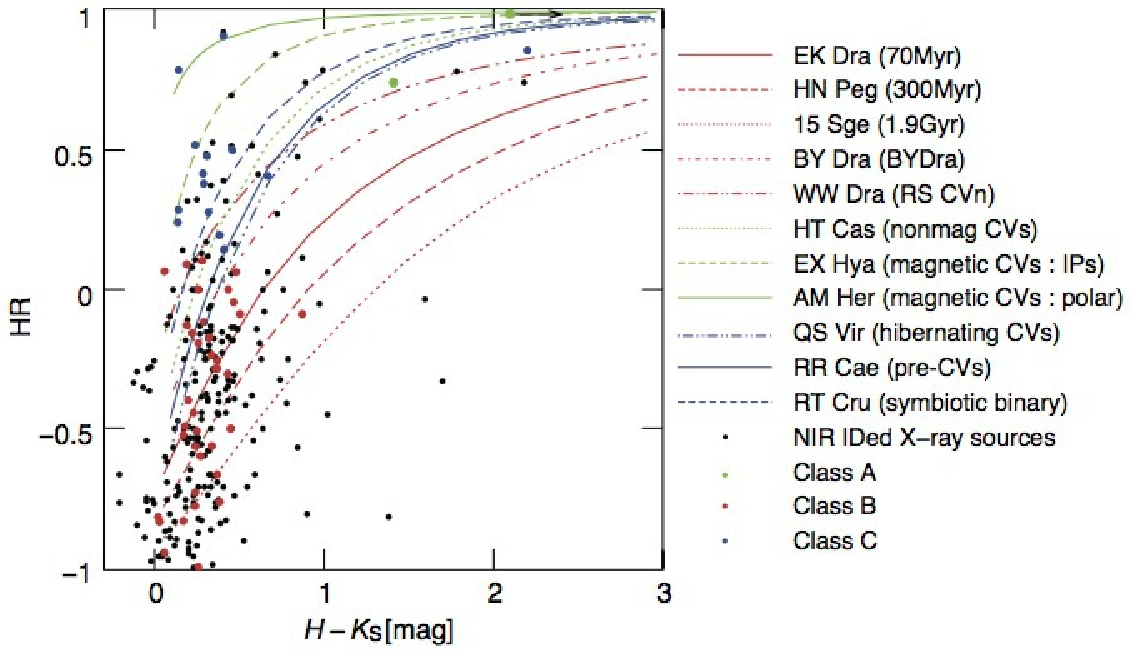}
 \end{center}
 \caption{Scatter plot of X-ray versus NIR colors (i.e., HR versus \hi--\ksi) for all
 the X-ray sources identified in the NIR imaging studies (small dots). For a source with
 no \hi-band detection, a black arrow is shown.
 Those with NIR spectra are shown with large circles with different colors for different classes. The
 extinction curves are calculated for some representative sources by assuming an
 intrinsic NIR color \citep{Cox00} and X-ray spectrum from the following work:
 (1) X-ray active single single stars with different ages (EK Dra: 70 Myr, HN Peg: 300 Myr, and 15 Sge: 1.9 Gyr; \cite{Gudel99}), 
 (2) X-ray active binary stars (BY Dra; \cite{Dempsey97}, WW Dra ; \cite{Dempsey93}),
 (3) a non-magnetic CV (HT Cas ; \cite{Nucita09}),
 (4) an intermediate polar (EX Hya; \cite{Allan98}), 
 (5) a polar (AM Her; \cite{Ishida97}),
 (6) hibernating CVs (QS Vir ; \cite{Matranga12}), (7) pre-cataclysmic variables (RR Cae ; \cite{Bilikov10}), 
 and (8) symbiotic bianries (RT Cru; \cite{Luna07}). 
 We calculate the expected \hi--\ksi~using
 \nh = 3.5$\times$10$^{22}$$\times$\textit{E (\hi--\ksi)} cm$^{-2}$ ~\citep{Nebot13}.
  }
 \label{f08}
\end{figure}

We critically examine the possibility that the class C sources are not intrinsically hard X-ray 
sources but only apparently hard
due to larger Galactic extinctions than class B sources that share the
same NIR characteristics as class C sources. 
We investigated individual sources by
plotting X-ray hardness (HR) versus infrared color, \hi--\ksi~(figure~\ref{f08}). As the
extinction increases, the \hi--\ksi~and the HR values become larger along the
extinction curve. The extinction curves are shown for some representative sources of CVs
(for class A) and X-ray active stars (for class B). It is found that class C sources are
not on the extinction curve of class B sources. Therefore, 
class C sources are likely to represent a new class distinctive from class A and B.

\subsection{Nature of the New Class}\label{s4-2}
What are the new class C sources?  We calculated the X-ray to NIR flux ratio (absorption
not corrected) for all the sources in tables~\ref{t04} and \ref{t05}. We used the hard
band for the X-ray and \textit{K}$_{\mathrm{s}}$ band for the NIR flux to minimize the
effect of interstellar attenuation. The median of the \fx/\fks\ values are
9.4$\times$10$^{-3}$ (class A), 7.8$\times$10$^{-4}$ (class B), and 8.4$\times$10$^{-3}$
(class C), while the rms of the values are 1.3$\times$10$^{-3}$ (class A),
1.3$\times$10$^{-4}$ (class B), and 2.2$\times$10$^{-3}$ (class C).

We estimated \fx/\fks~of CVs using \citet{baskil05} and \citet{ezuka99}, and found that their values are always larger
than \fx/\fks =1.35$\times$10$^{-3}$.
We also investigated \fx/\fks~of X-ray active stars in \citet{Pandey08} and  \citet{Pandey12}.
In contrast to CVs,  only 2 of 10 sources have \fx/\fks~values  greater than $\sim$10$^{-3}$ (the 
highest is 5.5$\times$10$^{-3}$) and the rest have much smaller values of \fx/\fks~between  10$^{-6}$ and 10$^{-4}$.  
Comparing these values with the average \fx/\fks~values of class A, B and C sources, the class A and B are most likely 
to be CVs and X-ray active stars, respectively. Most class C sources are presumably systems holding  white dwarfs
like CVs,  but may include some extremely X-ray active stars.

As mentioned above, class C has the following characteristics as a group: (1)  large \fx/\fks~values as those for CVs. (2) NIR spectra show 
no emission lines, but only absorption features of cool stars. (3) Hard X-ray spectra showing a possible Fe K feature (see figure~\ref{f07}). 
The fact (1) suggests degenerate system with low accretion rates. 
From (2), class C sources are probably not CVs, since CVs show emission lines 
such as H\emissiontype{I} (Br$\gamma$), He\emissiontype{I}, and He\emissiontype{II} 
(e.g., \cite{dhillon95, dhillon97}). From (3), active stars and binaries, which show relatively 
soft X-ray energy spectra,  are probably ruled out. 
Also, we consider that the majority of our class C sources are not HMXBs either,  that are known not to be the dominant population of hard X-ray sources (e.g., \cite{Laycock05}).


Therefore, class C are likely to be other binary systems holding degenerate stars.
Some binaries including white dwarfs, such as QS Vir and V471 Tau, are known to have similar properties 
to class C, namely,  strong Fe K emission line in X-rays with low X-ray luminosities of \lx~=
10$^{28-29}$~erg~s$^{-1}$ (e.g., \cite{Matranga12}, \cite{Bilikova10}) and the Na\emissiontype{I}, Ca\emissiontype{I}, 
and CO absorption lines in NIR (e.g., \cite{Howell10}). 

In fact, there are several such populations to exhibit the characteristics of class C that 
are hibernating CVs, pre-Cataclysmic Variables (pre-CVs),  and symbiotic binaries. 
Hibernating CVs are temporarily or marginally detached binary systems hosting a white dwarf and a dwarf star
where no or little accretion takes place. Pre-CVs are detached binary systems hosting a white dwarf and a late-type star (mainly M
-dwarf) with low level accretion, as opposed to the conventional CVs that are semi-detached systems with high level accretion.
Since both populations have been poorly studied,  it is conceivable that there might be a lot of undiscovered hibernating CVs and 
pre-CVs in our Galaxy. Recently, the number of white dwarf binaries, including detached systems, has increased significantly 
by large surveys such as the Sloan Digital Sky Survey and follow-up observations being likely to continue to grow in future
 (\cite{Ritter03} and its online updates at \url{http://wwwmpa.mpa-garching.mpg.de/RKcat/}).
Alternatively, symbiotic binaries, which are interacting binaries consisting of a mass-losing red giant and a 
blue companion (mainly white dwarf), is another candidate population of the new class. Symbiotic binaries 
are more luminous in NIR as they host red giants. Although many symbiotic binaries emit soft X-rays, a small 
fraction emit hard X-rays with $\sim$10 keV plasma temperature (e.g., \cite{Smith08}). They show strong Fe
features in X-rays~(e.g.,\cite{Smith08, Eze14}) and do not show Br$\gamma$ emission lines in NIR 
\citep{Schmidt03}. It is worth mentioning that some symbiotic binaries are actually discovered in the follow-up 
studies of newly discovered X-ray sources \citep{Berg06}.

To compare these possible populations with class C sources, we added
extinction curves of a hibernating CV, a pre-CV, and a symbiotic star in figure~\ref{f08}.
We see that both the possible populations and the class C sources scatter between class A and B sources  as a group. 
This suggests that at least some of the class C sources are either hibernating CVs, pre-CVs or symbiotic stars, having 
the intermediate HR between CVs and late-type stars.

\subsection{Contributions to the GRXE}\label{s4-3}
\begin{figure*}[htbp]
 \begin{center}
  \includegraphics[width=160mm]{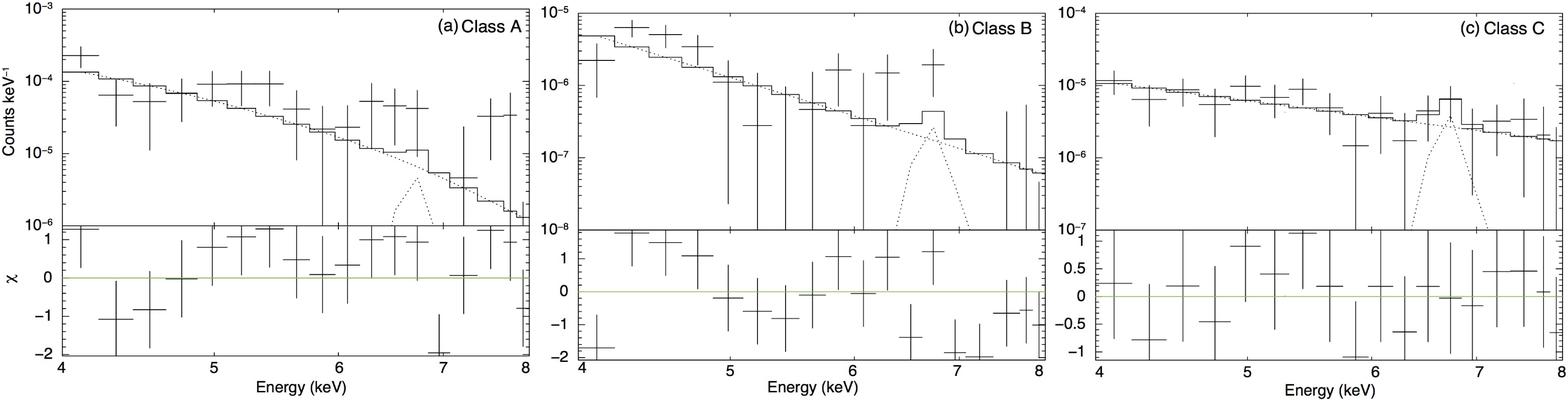}
  \caption{Fe K feature in the composite spectra of the class (a) A, (b) B, and (c) C
  sources. The best-fit parameters are shown in table~\ref{t07}.}\label{f09}
 \end{center}
\end{figure*}

How much fractional contribution does each class make for the GRXE, especially in the Fe
K band? To answer this question, we fitted the composite spectra in the 4--8~keV band
phenomenologically using a power-law continuum plus a Gaussian line representing the Fe
K-band emission. The free parameters are power-law index, normalization, the Gaussian
line width, and normalization. The Gaussian line center was fixed at 6.7 keV. The result
(table~\ref{t07}, figure~\ref{f09}) indicates that the class C sources contribute
$\sim$56\% of the Fe K-band emission of all the identified sources. 
It is thus possible that class C sources make an important contribution to the GRXE, especially at 
the Fe K band, however this cannot be confirmed with the data at hand, since the sources we selected 
for the NIR study are not uniformly distributed among the three classes.

\begin{table}[htbp]
 \caption{Best-fit parameters for spectral fittings in 4--8 keV.}\label{t07}
\begin{center}
 \begin{tabular}{lcc}
  \hline
   \hline
   Class  & $F_{\mathrm{Gau}}$\footnotemark[$*$]   & $\chi^{2}$/d.o.f \\
              & ($\ergcms$)  &                          \\
   \hline
   A    &  1.92$_{-0.41}^{+0.63}\times10^{-16}$  &  \phantom{-}17.16/14\\
   B   &   1.60$_{-0.38}^{+0.52}\times10^{-17}$  &  \phantom{-}23.68/15\\
   C   &   2.61$_{-0.25}^{+0.46}\times10^{-16}$   &  \phantom{-}\phantom{-}6.14/15\\
\hline
\end{tabular}
\par \noindent
\footnotemark[$*$] Flux of the gaussian component.
\end{center}
\end{table}

\section{Summary}\label{s5}
We presented results of a NIR spectroscopic follow-up study of dim X-ray point
sources constituting the GRXE in the two fields of the GP. The main results are
summarized as follows.

\begin{enumerate}
 \item Well-exposed \ksi-band spectra were obtained from a total of 65 X-ray sources
       in the two fields studied by Chandra deep exposures.
 \item Based on the NIR spectroscopic features and X-ray colors, we divided the sources into
       three categories: 
        (A) hard X-ray sources with emission features in NIR such as 
       H\emissiontype{I}, He\emissiontype{I}, and H$_{\rm II}$, 
       (B) soft X-ray sources with absorption features in NIR such as H\emissiontype{I}, Na\emissiontype{I},
       Ca\emissiontype{I}, and CO band heads, and 
       (C) hard X-ray sources with absorption features in NIR such as H\emissiontype{I}, Na\emissiontype{I},
       Ca\emissiontype{I}, and CO bandheads.
 \item We propose that class A and B sources are mainly comprised of CVs and X-ray
       active stars, whereas class C sources is presumably a different class distinctive from the
       two. We suggest that class C are binary systems hosting white dwarfs and late-type stars with low-level of accretion. 
       Possible candidates of the class C sources are hibernating CVs, pre-CVs and symbiotic stars.
 \item It is possible that this newly discovered class of sources contribute to a
       non-negligible fraction of the GRXE, especially in the Fe K band.
\end{enumerate}

\medskip

This work is based on data collected at Subaru Telescope operated by the National
Astronomical Observatory of Japan and on observations made with an ESO Telescope at the
La Silla Observatory under program ID 075.D-0767. IRAF is distributed by the National
Optical Astronomy Observatories, which are operated by the Association of Universities
for Research in Astronomy, Inc., under cooperative agreement with the National Science
Foundation. This research has made use of SAOImage DS9, developed by Smithsonian
Astrophysical Observatory.
K.\,M. is financially supported by the University of Tokyo global COE program, the
Hayakawa Satio Foundation by the Astronomical Society of Japan, the Sasagawa
Scientific Research Grant from The Japan Science Society, and  the Hyogo Science and Technology Association.
M.\,T is financially supported by MEXT/JSPS KAKENHI Grant Numbers 24105007 and 15H03642.

\bibliographystyle{aa}
\bibliography{ms}
\end{document}